\documentclass[pdflatex,sn-mathphys-num]{sn-jnl}

\usepackage{graphicx}%
\usepackage{multirow}%
\usepackage{amsmath,amssymb,amsfonts}%
\usepackage{amsthm}%
\usepackage{mathrsfs}%
\usepackage[title]{appendix}%
\usepackage{xcolor}%
\usepackage{textcomp}%
\usepackage{manyfoot}%
\usepackage{booktabs}%
\usepackage{algorithm}%
\usepackage{algorithmicx}%
\usepackage{algpseudocode}%
\usepackage{listings}%
\usepackage{amsmath}
\usepackage{array}         
\usepackage{booktabs}       
\usepackage{multirow}      
\usepackage{enumitem}
\usepackage{subcaption}
\usepackage{dsfont}

\theoremstyle{thmstyleone}%
\theoremstyle{thmstyletwo}%

\theoremstyle{thmstylethree}%

\begin{document}

\title[Article Title]{Explainable Attention-Guided Stacked Graph Neural Networks for Malware Detection}

\author[1]{\fnm{Hossein} \sur{Shokouhinejad}}

\author[1]{\fnm{Roozbeh} \sur{Razavi-Far}}

\author[1]{\fnm{Griffin} \sur{Higgins}}

\author[1]{\fnm{Ali A} \sur{Ghorbani}}

\affil*[1]{\orgdiv{University of New Brunswick}, \orgname{Faculty
 of Computer Science}, \orgaddress{\street{3 Bailey Dr.}, \city{Fredericton}, \postcode{E3B5A3}, \state{New Brunswick}, \country{Canada}}}

\abstract{Malware detection in modern computing environments demands models that are not only accurate but also interpretable and robust to evasive techniques. Graph neural networks (GNNs) have shown promise in this domain by modeling rich structural dependencies in graph-based program representations such as control flow graphs (CFGs). However, single-model approaches may suffer from limited generalization and lack interpretability, especially in high-stakes security applications. In this paper, we propose a novel stacking ensemble framework for graph-based malware detection and explanation. Our method dynamically extracts CFGs from portable executable (PE) files and encodes their basic blocks through a two-step embedding strategy. A set of diverse GNN base learners, each with a distinct message-passing mechanism, is used to capture complementary behavioral features. Their prediction outputs are aggregated by a meta-learner implemented as an attention-based multilayer perceptron, which both classifies malware instances and quantifies the contribution of each base model. To enhance explainability, we introduce an ensemble-aware post-hoc explanation technique that leverages edge-level importance scores generated by a GNN explainer and fuses them using the learned attention weights. This produces interpretable, model-agnostic explanations aligned with the final ensemble decision. Experimental results demonstrate that our framework improves classification performance while providing insightful interpretations of malware behavior.}


\maketitle

\section{Introduction}
Malware, or malicious software, continues to pose a significant threat to modern computing environments, particularly in enterprise systems, critical infrastructure, and cloud platforms. As malware evolves in complexity and evasion techniques, conventional signature-based detection methods struggle to keep pace. To address these limitations, researchers have turned to machine learning and deep learning techniques, which offer the potential to detect novel or obfuscated malware through behavioral and structural analysis~\cite{Our_Survey}. However, the effectiveness of such models largely depends on the quality of features and the representation of program behavior, necessitating advanced techniques capable of capturing the intricate dependencies within program execution.

Graph Neural Networks (GNNs) have recently emerged as powerful tools for learning from graph-structured data and have demonstrated considerable success in a range of cybersecurity tasks, particularly malware detection~\cite{Consistency,MalGNE}. Unlike traditional neural networks, GNNs are specifically designed to operate on graph data by iteratively aggregating information from neighboring nodes and edges, allowing them to capture intricate structural and relational dependencies across the graph. This capability makes GNNs especially well-suited for modeling software behavior, which is often naturally represented as a graph~\cite{NCP}. In malware detection, GNNs have been applied to various graph-based representations of programs, including function call graphs (FCGs)~\cite{FCG_1}, API call graphs~\cite{API_Call_Graph_1}, and most importantly, control flow graphs (CFGs)~\cite{CFG_1}, each of which captures different behavioral characteristics.
Among these representations, CFGs are particularly valuable because they illustrate the execution flow of a program in a structured and analyzable form. A CFG models basic blocks of instructions as nodes and the possible control transfers between them as directed edges. This abstraction captures both high-level logic and low-level execution patterns, enabling the identification of structural anomalies and irregular control paths that are often indicative of malicious behavior. When combined with GNNs, CFGs enable the extraction of rich representations that capture both local execution behaviors and global structural context, thereby improving the detection of sophisticated and evasive malware.

While individual GNN models have shown promising performance, leveraging ensemble learning can further improve generalization, robustness, and predictive performance~\cite{9_aaai.v37i7.26000,10_ijcai2022p289,11_10.5555/3666122.3668333}. Ensemble techniques combine the outputs of multiple base learners to reduce variance and avoid overfitting~\cite{12_ZENG2025106870}. Among these methods, stacking stands out as a powerful strategy that trains a meta-learner to combine the predictions of multiple base GNN models. Each GNN in the ensemble may capture different aspects of the input graph due to architectural or training diversity, and the meta-learner can synthesize these diverse outputs into a more accurate and reliable prediction. This hierarchical learning paradigm has been underexplored in malware detection with GNNs and offers a promising direction for improving detection accuracy and resilience against evasive techniques~\cite{2_malware_stacking,5_NAEEM2023119952,6_9099045,7_9165209}.

In addition to achieving high classification performance, the explainability of malware detection models has become a critical research focus, particularly for applications in high-assurance systems where transparency and trust are essential~\cite{Dual,Hesam}. Understanding which parts of the CFG contribute most to a detection decision not only helps validate the model’s outputs but also provides actionable insights for security analysts and incident responders. GNNs, while effective in modeling structural patterns in program behavior, often lack interpretability due to their complex architecture. To address this, eXplainable Artificial Intelligence (XAI) techniques such as GNNExplainer~\cite{GNNExplainer}, PGExplainer~\cite{PGExplainer}, SubgraphX~\cite{SubgraphX}, and CaptumExplainer~\cite{SAL_GBP} have been proposed to highlight influential nodes, edges, or subgraphs that drive predictions. When applied to a stacking ensemble (SE), explanation becomes more challenging, as the meta-learner integrates outputs from multiple base models, each potentially capturing different aspects of the graph.

In this study, we address the challenge of graph-based malware detection and interpretability through a SE learning framework. The approach begins with the dynamic extraction of CFGs from Portable Executable (PE) files. Each basic block within the CFG is then embedded using a two-step feature representation process. For the classification task, we employ a GNN-based SE in which the base learners consist of diverse GNN models, each utilizing a distinct message-passing strategy to ensure representational diversity. The prediction outputs of these base models serve as inputs to a meta-learner, implemented as a multi-layer perceptron (MLP) enhanced with an attention mechanism. This meta-learner not only performs the final classification but also produces attention scores that reflect the relative contribution of each base learner to the final decision. For model explainability, we adopt a post-hoc explanation approach by integrating a state-of-the-art GNN explainer to assign importance scores to nodes and edges within each base model. These individual explanations are then aggregated using the attention scores from the meta-learner, leading to the development of a novel explanation method tailored for the stacking framework. The main contributions of this study are as follows:
\begin{itemize}
    \item We propose a novel SE framework for graph-based malware detection, where diverse GNN base learners capture complementary structural and semantic information from control flow graphs extracted from PE files.
    \item A meta-learner equipped with an attention mechanism is introduced to aggregate predictions from base learners, enabling both accurate classification and interpretable attribution of each model’s contribution to the final decision.
    \item We develop a new post-hoc explanation method that combines state-of-the-art GNN explanation techniques with the attention scores from the meta-learner to generate enhanced, ensemble-aware interpretations of malware predictions.
\end{itemize}

The remainder of this paper is structured as follows: Section~\ref{sec:related_works} reviews prior work on GNN-based malware detection, ensemble learning, and explainable GNNs. Section~\ref{sec:framework} presents the proposed SE framework, including the dynamic CFG extraction process, node feature embedding, base learner architecture, meta-learner design, and the aggregation-based explanation method. Section~\ref{sec:result} reports the experimental setup, datasets, performance evaluation, and explainability analysis. Finally, Section~\ref{sec:conclusion} concludes the paper and discusses directions for future work.

\section{Related Works}\label{sec:related_works}
Recent research has increasingly focused on advancing malware detection through graph-based learning techniques, particularly those that model program behavior using CFGs. GNNs have shown strong performance in capturing the structural dependencies within such graphs, enabling good classification of malicious software. At the same time, ensemble learning approaches, particularly stacking-based frameworks, have been explored to combine the strengths of multiple models for improved robustness and accuracy. As these models grow in complexity, the need for explainability has become more pronounced, especially for deployment in security-critical systems. This section reviews prior work across three major areas: GNN-based malware detection, stacking-based ensemble methods for malware analysis, and explainable GNN frameworks for interpreting malicious behavior.

Several studies have leveraged GNNs for malware detection using CFGs as input representations. Peng et al.~\cite{MalGNE} introduced MalGNE, a node embedding framework that begins by encoding assembly instructions using a rule-based vectorization scheme to handle the out-of-vocabulary issue. It employs aggregation and attention-based bidirectional LSTM layers to capture the sequential and semantic structure of instructions within basic blocks. These representations are then passed to a GNN for classification. In a related effort, Zhang et al.~\cite{CFG_1} proposed a few-shot malware classification model based on a triplet-trained graph transformer. Their method processes each malware sample as a CFG and learns embeddings that capture structural and semantic relationships among basic blocks. Using triplet loss, the model is trained to place similar samples closer in the embedding space while distancing dissimilar ones, thereby improving generalization in data-scarce scenarios.

Further advancing this direction, Amer et al. \cite{new_2} proposed GraphShield, a dynamic graph-based malware detection framework that constructs temporal graphs from API call sequences and leverages attention-enhanced GNN architectures for behavioral analysis. GraphShield integrates explainability through GNN interpretation tools that highlight key API subgraphs responsible for malicious behavior, demonstrating substantial gains in accuracy and interpretability compared to sequential deep models. Similarly, Wang et al. \cite{new_3} introduced SIGDroid, a knowledge-distillation-driven Android malware detection framework that transforms heterogeneous FCGs into homogeneous graphs while preserving semantic and permission-related features. Their student-centered distillation approach enables compact GNNs to achieve robust performance across large-scale datasets, addressing model scalability and heterogeneity issues in FCG-based learning.In the context of securing critical infrastructures, Esmaeili et al. proposed a GNN-based adversarial IoT malware detection framework that detects manipulated CFGs before classification. Their anomaly detection approach effectively identifies adversarially perturbed graphs from campaigns such as Gafgyt, Mirai, and Tsunami, achieving over 98\% detection accuracy. This study underscores the importance of adversarial robustness and demonstrates the viability of CFG-based GNNs in protecting industrial and IoT environments \cite{new_1}.

In addition to GNN-based methods, several studies have employed SE techniques to enhance detection performance and address challenges such as data imbalance and feature redundancy. Li et al.~\cite{2_malware_stacking} proposed a hybrid detection approach that combines information gain and principal component analysis (IG-PCA) for feature selection, followed by an SE framework. The model uses an attention-based meta-learner to integrate diverse base classifiers and adaptively weight their contributions. Similarly, Naeem et al.~\cite{5_NAEEM2023119952} presented a malware detection technique that transforms memory dumps into images and extracts handcrafted features, which are then classified using an ensemble of convolutional neural networks. A fully connected neural network serves as the meta-learner, enhancing the final decision. Another example is SEDMDroid~\cite{6_9099045}, which combines static and dynamic features (such as permissions, API calls, network behavior, and system activity) using an SE of decision trees, support vector machines, and random forests, with a neural network acting as the meta-classifier. Vasan et al.~\cite{7_9165209} further explored ensemble learning for cross-architecture malware detection on IoT devices. Their method extracts opcode sequences from binaries compiled for various architectures, converts them into images, and processes them using deep learning models. The final predictions are aggregated using a fully connected neural network to improve detection performance across platforms.

As model interpretability becomes increasingly important, recent studies have proposed explanation techniques tailored for GNN-based malware classifiers. Herath et al.~\cite{CFGExplainer} presented a model-agnostic framework that identifies influential subgraphs within CFGs and ranks node importance using a surrogate model trained on node embeddings. This approach helps visualize which parts of the graph contribute most to classification. Building on this line of work, the study in~\cite{Consistency} introduced a dynamic CFG-based malware detection framework that employs a hybrid node embedding method combining rule-based encoding with autoencoder-derived features. After classification via a GNN, the framework applies various explanation techniques, including GNNExplainer, PGExplainer, and Captum with different attribution strategies. The authors also proposed a new explanation method, RankFusion, which aggregates scores from multiple explainers to generate more stable and informative attributions.

\begin{figure}
    \centering
    \includegraphics[width=\linewidth]{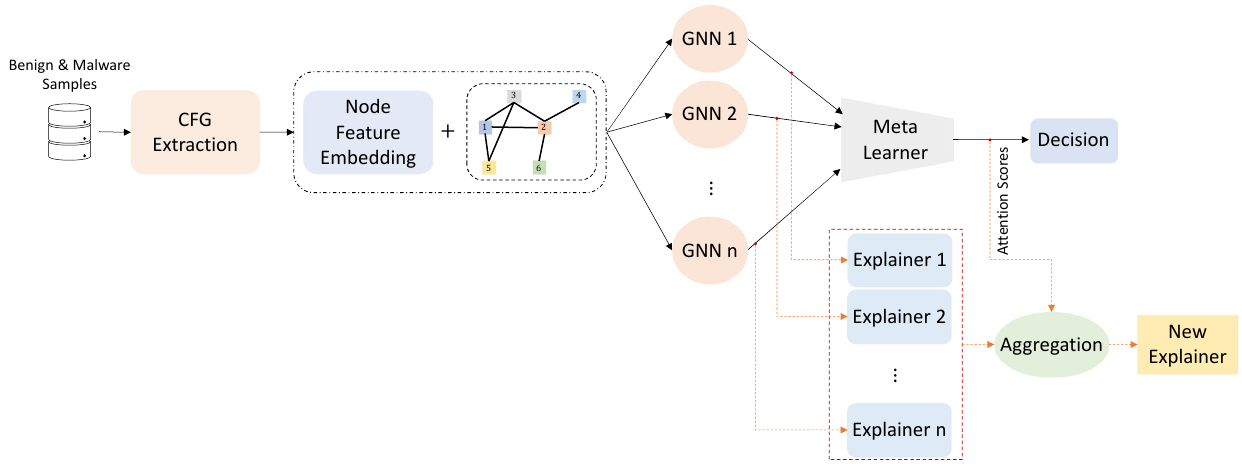}
    \caption{Overall architecture of the proposed explainable malware detection method.}
    \label{fig:end2end}
\end{figure}

\section{Proposed Method}\label{sec:framework}

\begin{figure}
    \centering
    \includegraphics[width=\linewidth]{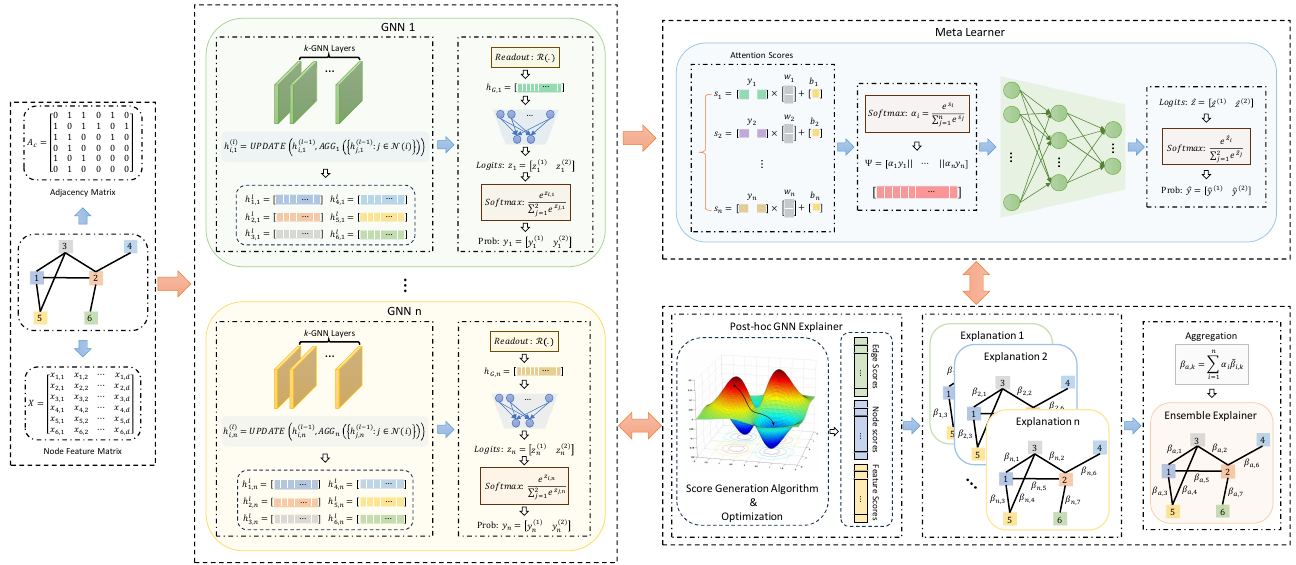}
    \caption{Detailed framework for explainable malware detection using a stacking ensemble of GNNs.}
    \label{fig:framework}
\end{figure}

This section presents our proposed framework for explainable malware detection using a GNN-based SE approach. An overview of the proposed method is illustrated in Figure~\ref{fig:end2end}, providing a concise visual summary of the overall process. The detailed architecture of the GNN-based SE framework is subsequently illustrated in Figure~\ref{fig:framework}. Our method consists of four key components. First, CFGs are dynamically extracted from PE files, and a two-step embedding strategy is applied to encode semantic and structural features of each basic block. Second, multiple GNN models with distinct message-passing mechanisms are used as base learners to capture complementary aspects of the graph data. Third, the outputs of these base learners are fed into an attention-enhanced MLP that serves as the meta-learner, responsible for final classification and generating model-level attribution scores. Finally, we introduce a novel post-hoc explainability method that aggregates the edge-level explanation from individual base models, weighted by the attention scores produced by the meta-learner, to produce ensemble-aware interpretations. In essence, the explanation reveals which patterns are consistently emphasized across the ensemble when identifying malware behavior. The following subsections provide a detailed description of each component.
\subsection{Dynamic CFG Extraction and Node Feature Embedding}
CFGs model the execution flow of a program, where nodes represent basic blocks, which are sequential groups of instructions with a single entry and a single exit point. Edges indicate possible transitions between these blocks. Static CFGs, constructed through disassembly, often miss important execution paths due to obfuscation, indirect jumps, or dynamic code loading. In contrast, dynamic CFGs are generated by tracing the actual runtime behavior of a program, capturing execution paths that may not be visible through static analysis. This makes them more effective for analyzing advanced or evasive malware. In our framework, we extract dynamic CFGs from PE files to build accurate graph representations for downstream feature embedding and classification.

\begin{figure}
    \centering
    \includegraphics[width=\linewidth]{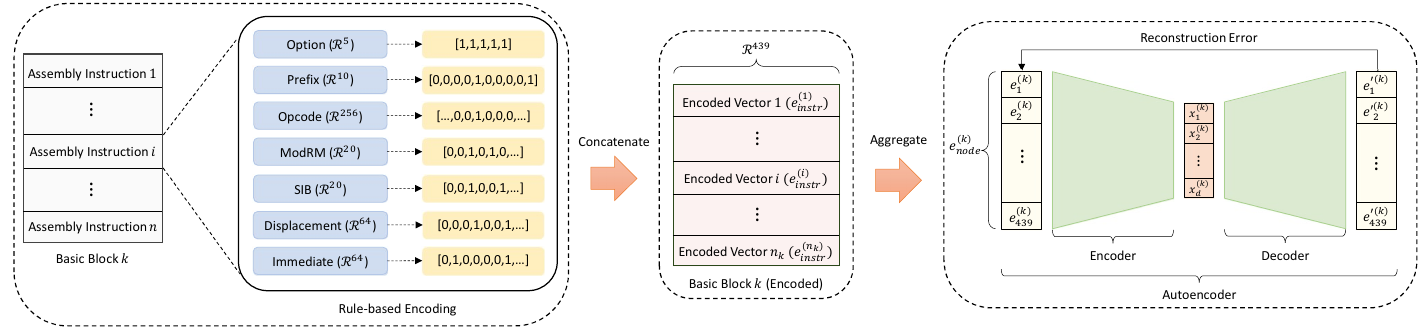}
    \caption{The two-step node feature embedding process, including rule-based instruction encoding and autoencoder-based dimensionality reduction.}
    \label{fig:embedding_pipeline}
\end{figure}

To prepare each node in the dynamically extracted CFG for downstream learning, we employ a two-step feature embedding strategy, illustrated in Figure~\ref{fig:embedding_pipeline}. The first step involves rule-based encoding of assembly instructions within each basic block, and the second step applies unsupervised dimensionality reduction using an autoencoder to obtain compact representations.

Each basic block consists of a sequence of x86-64 assembly instructions. Following a modified version of the approach proposed in~\cite{MalGNE}, in the first step, each individual instruction within a basic block is transformed into a fixed-length feature vector using a fine-grained, rule-based encoding scheme. This design provides a unique and full-coverage mapping for the x86-64 instruction set, thereby eliminating the out-of-vocabulary problem common in text-based embeddings. Specifically, each instruction is decomposed into up to seven components: prefix, opcode, ModRM, SIB, displacement, immediate, and an option flag, and encoded as follows:
\begin{enumerate}
    \item \textbf{Prefix:} The prefix includes four fields: the extra segment (ES) register, operand-size override, address-size override, and lock prefix. The ES segment register has seven possible values, while the other three fields are binary (0 or 1). Thus, the prefix is encoded as a ten-dimensional one-hot vector.
    \item \textbf{Opcode:} The opcode specifies the fundamental operation carried out by the instruction and can assume up to 256 distinct values. It is represented using a 256-dimensional one-hot encoding.
    \item \textbf{ModRM:} The ModRM field is a single byte that is partitioned into three segments: two bits for the mode, three bits for the register, and three bits for the memory address. These correspond to four, eight, and eight possible values, respectively. As a result, the ModRM field is represented as a 20-dimensional one-hot vector.
    \item \textbf{SIB:} The Scale-Index-Base (SIB) field is also a single byte, partitioned into three segments: two bits for the scale factor, three bits for the index, and three bits for the base register. These correspond to four, eight, and eight possible values, respectively, resulting in a 20-dimensional one-hot encoding for the SIB field.
    \item \textbf{Displacement:} In x86-64 assembly, the displacement represents an offset value employed in memory addressing to compute the effective address. It forms part of the instruction’s addressing mode and is added to a base or index register to obtain the final memory location. The displacement is encoded as a 64-dimensional binary vector.
    \item \textbf{Immediate:} The immediate value is a constant that is directly embedded within the instruction. Unlike the displacement, which contributes to memory addressing, the immediate serves as a direct operand in the execution of the instruction. It is encoded as a 64-dimensional binary vector.
    \item \textbf{Option:} Because the prefix, ModRM, SIB, displacement, and immediate fields may or may not appear in an instruction, a 5-dimensional binary vector is employed to indicate their presence or absence.
\end{enumerate}

The concatenation of these encodings produces a 439-dimensional feature vector for each instruction, ensuring that both the control semantics and operand structures are preserved.

Since a basic block may contain multiple instructions, we apply an aggregation function over their encoded vectors to produce a unified high-dimensional representation for the node:
\begin{equation}
{e}_{\text{node}}^{(k)} = \text{AGG}\left( {e}_{\text{instr}}^{(1)}, {e}_{\text{instr}}^{(2)}, \ldots, {e}_{\text{instr}}^{(n_k)} \right)
\end{equation}
where \( n_k \) is the number of instructions in the \(k\)-th node (basic block), \( {e}_{\text{instr}}^{(i)} \in \mathbb{R}^{439} \) is the encoded vector of the \(i\)-th instruction, and \( \text{AGG}(\cdot) \) denotes an aggregation function such as mean or max pooling. The result \( {e}_{\text{node}}^{(k)} \in \mathbb{R}^{439} \) is the aggregated high-dimensional representation of node \(k\).

To reduce the dimensionality of these vectors while preserving essential information, we train an autoencoder in an unsupervised manner. The autoencoder consists of an encoder-decoder architecture and is optimized using the mean squared error (MSE) loss:
\begin{equation}
L_{\text{MSE}} = \frac{1}{M} \sum_{i=1}^{M} \left\| {e}_{\text{instr}}^{(i)} - g_{\phi}(f_{\theta}({e}_{\text{instr}}^{(i)})) \right\|^2
\end{equation}
where \( f_{\theta} \) and \( g_{\phi} \) denote the encoder and decoder functions, respectively, and \( M \) is the number of instruction samples used for training.

After training, the encoder maps each node’s high-dimensional vector into a lower-dimensional latent space:
\begin{equation}
{x}^{(k)} = f_{\theta}({e}_{\text{node}}^{(k)})
\end{equation}
where \( {x}^{(k)} \in \mathbb{R}^{d} \), with \( d \ll 439 \), represents the compact embedding of the \(k\)-th node.

This two-step embedding process enables the generation of expressive and compact node features, which are well-suited for downstream graph learning tasks, particularly in settings where node-level supervision is unavailable.

By collecting the embeddings of all nodes, we construct the node feature matrix \( {X} \in \mathbb{R}^{N \times d} \), where \( N \) is the total number of nodes in the graph. Each row of this matrix corresponds to one node's embedding:
\begin{equation}
{X} = 
\begin{bmatrix}
{x}^{(1)} \\
{x}^{(2)} \\
\vdots \\
{x}^{(N)}
\end{bmatrix}
\in \mathbb{R}^{N \times d}
\end{equation}

\subsection{Graph Neural Networks as Base Learners}
GNNs are a class of neural models designed to learn from graph-structured data. In our setting, the input is a CFG in which each node represents a basic block with an associated feature vector \( {x}^{(i)} \in \mathbb{R}^d \), and edges reflect control flow relationships. GNNs operate through a process known as message passing, in which each node iteratively updates its representation by aggregating information from its neighbors.

The general message passing framework for a GNN layer is expressed as:
\begin{equation}
{h}_i^{(l)} = \text{UPDATE}^{(l)}\left({h}_i^{(l-1)}, \text{AGG}^{(l)}\left(\left\{ {h}_j^{(l-1)} : j \in \mathcal{N}(i) \right\}\right)\right)
\end{equation}
where \( {h}_i^{(l)} \) denotes the representation of node \( i \) at layer \( l \), \( \mathcal{N}(i) \) denotes the set of neighboring nodes of \( i \), and \( \text{AGG}(\cdot) \) and \( \text{UPDATE}(\cdot) \) are differentiable functions responsible for aggregating and updating node features. At the initial layer, \( {h}_i^{(0)} = {x}^{(i)} \), the input node feature vector.

In this work, we leverage multiple types of GNN models as base learners to introduce representational diversity, which is essential for the effectiveness of the SE. Different GNN architectures adopt distinct aggregation and update strategies, resulting in varied learning biases and node embeddings. By combining these diverse models, the meta-learner can learn complementary patterns that improve malware classification performance.

Among the many GNN variants proposed in the literature, three widely recognized and influential models are the Graph Convolutional Network (GCN)~\cite{GCN}, Graph Isomorphism Network (GIN)~\cite{GIN}, and Graph Attention Network (GAT)~\cite{GAT}. These models serve as prominent examples, each characterized by a distinct message passing mechanism, as outlined below.

\paragraph{Graph Convolutional Network (GCN)}  
GCN applies a normalized aggregation of neighboring features followed by a linear transformation:
\begin{equation}
{h}_i^{(l)} = \sigma\left( \sum_{j \in \mathcal{N}(i) \cup \{i\}} \frac{1}{\sqrt{d_i d_j}} {W}^{(l)} {h}_j^{(l-1)} \right)
\end{equation}
where \( {h}_i^{(l)} \) is the embedding of node \( i \) at layer \( l \), \( d_i \) and \( d_j \) denote the degrees of nodes \( i \) and \( j \), respectively, \( {W}^{(l)} \) is the learnable weight matrix at layer \( l \), and \( \sigma(\cdot) \) is a non-linear activation function such as ReLU.

\paragraph{Graph Isomorphism Network (GIN)}  
GIN is designed for high discriminative power and follows an MLP-based aggregation scheme:
\begin{equation}
{h}_v^{(l)} = \text{MLP}^{(l)}\left((1 + \epsilon) \cdot {h}_v^{(l-1)} + \sum_{u \in \mathcal{N}(v)} {h}_u^{(l-1)}\right)
\end{equation}
where \( {h}_v^{(l)} \) is the embedding of node \( v \) at layer \( l \), \( u \in \mathcal{N}(v) \) refers to the neighbors of node \( v \), \( \epsilon \) is either a learnable parameter or a fixed scalar, and \( \text{MLP}^{(l)}(\cdot) \) denotes a multi-layer perceptron. The use of summation ensures injective aggregation, making GIN highly expressive in distinguishing graph structures.

\paragraph{Graph Attention Network (GAT)}  
GAT introduces attention mechanisms to learn the importance of neighboring nodes:
\begin{equation}
{h}_v^{(l)} = \sigma\left( \sum_{u \in \mathcal{N}(v)} \alpha_{vu}^{(l)} {W}^{(l)} {h}_u^{(l-1)} \right)
\end{equation}
where \( \alpha_{vu}^{(l)} \) is the attention coefficient between node \( v \) and neighbor \( u \), computed based on their features, and \( {W}^{(l)} \) is a learnable weight matrix. The attention mechanism enables the network to assign varying influence to different neighbors during aggregation.

After several rounds of message passing, node embeddings capture local structural and feature information. To perform graph classification, a readout function is applied to aggregate node-level embeddings into a single graph-level vector. Common readout strategies include global mean pooling, max pooling, sum pooling, and more advanced techniques such as Set2Set or attention-based pooling. The resulting graph embedding, denoted as \( {h}_{G} \), serves as a compact representation of the entire graph.

This graph-level representation is then fed into a classifier, which is typically implemented as a simple MLP. For binary classification tasks such as malware detection, the final layer of the MLP outputs a two-dimensional logit vector \( {z} = [z^{(1)}, z^{(2)}] \in \mathbb{R}^2 \), where each component corresponds to one of the two classes (benign or malicious). A softmax function is subsequently applied to convert these logits into a probability distribution \( {y} = [y^{(1)}, y^{(2)}] \), where \( y^{(1)} \) and \( y^{(2)} \) indicate the predicted probabilities for the benign and malicious classes, respectively. The final predicted label corresponds to the class with the higher probability.

Using different GNN architectures as base learners enriches the feature space available to the meta-learner. Each GNN type captures distinct structural and semantic patterns due to differences in aggregation functions, update mechanisms, and learning biases. For malware detection, where subtle behavioral variations in control flow graphs can be indicative of malicious activity, such architectural diversity enhances classification performance. The meta-learner can leverage these complementary perspectives to improve generalization and decision accuracy.

\subsection{Attention-Based Stacking Ensemble Architecture}
\begin{figure*}
    \centering
    \includegraphics[width=\linewidth]{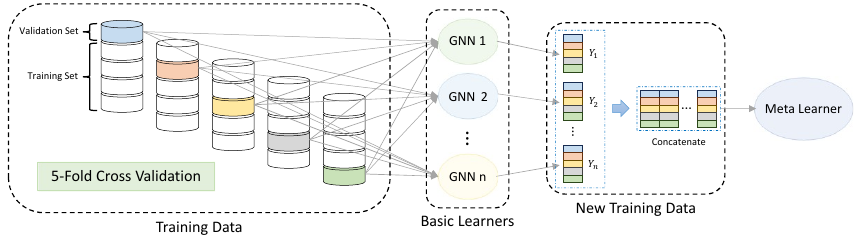}
    \caption{Training procedure of the attention-based SE model}
    \label{fig:training}
\end{figure*}

\begin{algorithm}[t]
\caption{Training the Attention-Based SE model.}
\label{alg:train}
\begin{algorithmic}[1]
\State \textbf{Input:} PE Training Dataset $\mathcal{D}_{\text{train}}$
\State \textbf{Output:} Base learners ($\text{GNN}_i$) and meta-learner (attention weights + MLP)

\Function{EncodeFeatures}{$\text{PE}$}
    \State Extract CFG $G=(V,E)$
    \State $X \gets [\,]$
    \For{$k \in V$}
        \State Encode instructions to ${e}_{\text{instr}}^{(t)} \in \mathbb{R}^{439}$
        \State ${e}_{\text{node}}^{(k)} \gets \text{AGG}({e}_{\text{instr}}^{(1)},\ldots,{e}_{\text{instr}}^{(n_k)})$
        \State ${x}^{(k)} \gets f_\theta({e}_{\text{node}}^{(k)}) \in \mathbb{R}^{d}$
        \State $X \gets [\,X;~ x^{(k)}\,]$ 
    \EndFor
    \State \Return $(G, X)$
\EndFunction

\For{each $(\text{PE},y)\in\mathcal{D}_{\text{train}}$}
    \State $(G,X) \gets$ \Call{EncodeFeatures}{$\text{PE}$}
\EndFor

\State Partition $\mathcal{D}_{\text{train}} \to \{\mathcal{D}^{(f)}\}_{f=1}^{5}$
\For{$i=1$ to $n$}
\State $Y_i \leftarrow [\,]$
    \For{$f=1$ to $5$}
        \State $\text{TrainSet} \gets \mathcal{D}_{\text{train}} \setminus \mathcal{D}^{(f)}$
        \State $\text{ValSet} \gets \mathcal{D}^{(f)}$
        \State Train $\text{GNN}_i^{(f)}$ on $(G, X, y) \in \mathcal{D}_{\text{TrainSet}}$
        \State $\widehat{Y}_i^{(f)} \gets \text{GNN}_i^{(f)}(\text{ValSet})$
        \State $Y_i \leftarrow \begin{bmatrix} Y_i ;~ \widehat{Y}_i^{(f)} \end{bmatrix}$
    \EndFor
\EndFor

\State $Y \gets [\,Y_1 \,\|\, Y_2 \,\|\, \cdots \,\|\, Y_n\,]$
\State Train meta-learner (attention weights $(w_i,b_i)$ and MLP) on $(Y, y)$

\For{$i=1$ to $n$}
    \State Train $\text{GNN}_i$ on $\{(G,X,y)\in\mathcal{D}_{\text{train}}\}$
\EndFor

\State \Return $\text{GNN}_i$, $(w_i,b_i)$, MLP
\end{algorithmic}
\end{algorithm}

Stacking ensemble (SE) learning is a powerful strategy for improving prediction accuracy and robustness by leveraging the complementary strengths of multiple models. Rather than relying on a single model’s output, stacking combines the predictions of several base learners through a secondary model, known as the meta-learner, to produce a more reliable final decision. This approach enhances generalization and reduces overfitting by mitigating individual model biases and variances.
In our work, we adopt an SE framework that integrates multiple GNN-based base learners with an attention-based meta-learner. The training procedure consists of three main phases: (i) independently training each base learner using cross-validation, (ii) generating a new meta-training dataset from the base learners’ validation predictions, and (iii) training the meta-learner using this aggregated information. The overall training procedure is illustrated in Figure~\ref{fig:training} and summarized in Algorithm~\ref{alg:train}.

Let \( \mathcal{D}_{\text{train}} \) denote the original training dataset, and suppose we use 5-fold cross-validation. The dataset is partitioned into 5 non-overlapping subsets \( \mathcal{D}^{(1)}, \ldots, \mathcal{D}^{(5)} \). Let \( \text{GNN}_i \) represent the \( i \)-th base learner. The following steps are applied to construct the training set for the meta-learner:

\begin{enumerate}
    \item \textbf{Fold-wise training:} For each base learner $\text{GNN}_i$, and each fold \( f = 1, \ldots, 5 \), a model is trained on training set \( \mathcal{D}_{\text{train}} \setminus \mathcal{D}^{(f)} \) and evaluated on validation set \( \mathcal{D}^{(f)} \).
    
    \item \textbf{Prediction on validation fold:} The trained model \( \text{GNN}_{i}^{(f)} \) produces predicted probability vectors for samples in \( \mathcal{D}^{(f)} \). Let:
    \begin{equation}
    \widehat{Y}_{i}^{(f)} = \text{GNN}_{i}^{(f)}\left(\mathcal{D}^{(f)}\right) \in \mathbb{R}^{|\mathcal{D}^{(f)}| \times 2}
    \end{equation}
    
    \item \textbf{Row-wise aggregation:} After all 5 folds, the predictions are stacked row-wise to form the complete prediction matrix for \( i \)-th base learner :
    \begin{equation}
    Y_i = 
    \begin{bmatrix}
    \widehat{Y}_i^{(1)} \\
    \widehat{Y}_i^{(2)} \\
    \vdots \\
    \widehat{Y}_i^{(5)}
    \end{bmatrix}
    \in \mathbb{R}^{|\mathcal{D}_{\text{train}}| \times 2}
      \end{equation}
   \end{enumerate}
 
After constructing \( \widehat{{Y}}_i \) for all \( n \) base learners, we concatenate them to form the input matrix for the meta-learner:
\begin{equation}
Y = \left[ Y_1 \, \| \, Y_2 \, \| \, \cdots \, \| \, Y_n \right] \in \mathbb{R}^{|\mathcal{D}_{\text{train}}| \times 2n}
\end{equation}
Each row of \( {Y} \) is the concatenation of predicted probabilities from all base learners for a single sample, and serves as input to the meta-learner.

After hyperparameter tuning and construction of the meta-training dataset through 5-fold cross-validation, each GNN base learner is retrained on the entire original training set \( \mathcal{D}_{\text{train}} \) (i.e., the union of all folds) to produce its final model. These final models are then used to generate predictions on the test set for evaluation by the meta-learner.

Let the input vector for a given sample be denoted as:
\begin{equation}
y = \left[ y_1 \, \| \, y_2 \, \| \, \cdots \, \| \, y_n \right] \in \mathbb{R}^{2n}
\end{equation}
where \( {y}_i \in \mathbb{R}^2 \) is the prediction vector from base learner \( \text{GNN}_i \).

Each \( {y}_i \) is passed through a learnable linear transformation to compute an attention score:
\begin{equation}
s_i = {w}_i^\top {y}_i + b_i, \quad \text{for } i = 1, \ldots, n
\end{equation}
where \( {w}_i \in \mathbb{R}^2 \) and \( b_i \in \mathbb{R} \) are learnable parameters. These scores are normalized via softmax to obtain attention weights:
\begin{equation}
\alpha_i = \frac{\exp(s_i)}{\sum_{j=1}^{n} \exp(s_j)}
\end{equation}

The final attention-aware input is constructed as:
\begin{equation}
\Psi = \left[ \alpha_1 {y}_1 \, \| \, \alpha_2 {y}_2 \, \| \, \cdots \, \| \, \alpha_n {y}_n \right] \in \mathbb{R}^{2n}
\end{equation}
This vector \( \Psi \) is then passed through an MLP, which outputs a two-dimensional logit vector \( \hat{{z}} \in \mathbb{R}^2 \), followed by a softmax operation to yield the predicted class probabilities \( \hat{{y}} \in \mathbb{R}^2 \).
The learned attention weights \( \alpha_i \) offer interpretability by indicating the relative contribution of each base learner to the ensemble’s final decision.

\subsection{Aggregation-Driven Explainability for Stacked GNNs}

\begin{figure*}
    \centering
    \includegraphics[width=\linewidth]{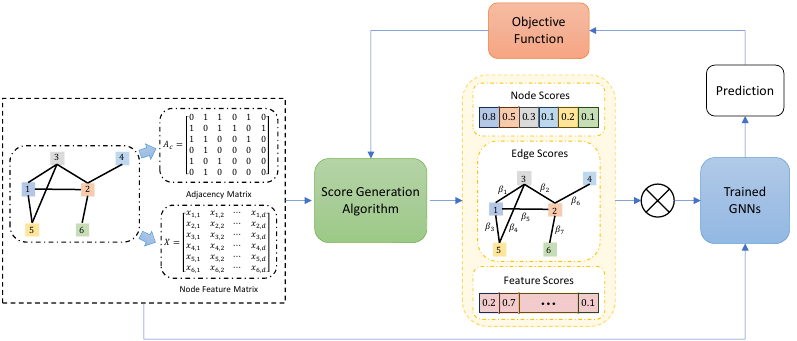}
    \caption{Post-hoc explainability process for interpreting GNN predictions.}
    \label{fig:explainer}
\end{figure*}

Post-hoc GNN explainability methods aim to interpret the predictions of trained GNN models by identifying important subgraphs in the input graph. These methods typically consist of three core components: (i) a pre-trained GNN model whose predictions are to be explained, (ii) a score generation mechanism that produces importance scores over graph elements (nodes, edges, or features), and (iii) an objective function that guides the explainer to identify input elements that most significantly influence the model’s prediction. A variety of explanation techniques have been proposed in the literature, including perturbation-based methods, gradient-based approaches, surrogate modeling techniques, and decomposition-based methods, each offering distinct advantages in terms of fidelity, interpretability, and computational cost. An overview of this process is illustrated in Figure~\ref{fig:explainer}.

In this work, we focus on edge-level explanations and utilize a post-hoc GNN explainer as base explainer to obtain importance scores for edges in each input graph. Let \( \beta_{i,k}\) denote the importance score assigned by the explainer to the \( k \)-th edge of a given sample as interpreted by the \( i \)-th base learner, where \( i = 1, \ldots, n \) and \( k = 1, \ldots, m \) (with \( n \) being the number of base learners and \( m \) the number of edges in the sample).

To ensure consistency across base learners, we first normalize the edge scores from each base model:
\begin{equation}
\widetilde{\beta}_{i,k} = \frac{\beta_{i,k}}{\sum_{j=1}^{m} \beta_{i,j}}
\end{equation}

Then, leveraging the attention weights \( \alpha_i \in [0,1] \) computed by the attention-based meta-learner, we aggregate the normalized edge scores into a unified importance score for each edge:
\begin{equation}
\beta_{a,k} = \sum_{i=1}^{n} \alpha_i \, \widetilde{\beta}_{i,k}, \quad \text{for } k = 1,\ldots,m
\end{equation}
Here, \( \beta_{a,k} \) represents the aggregated edge importance for the \( k \)-th edge, capturing both the local importance from base learners and their global contribution to the ensemble decision through \( \alpha_i \).

This aggregation-driven approach provides a unified and interpretable explanation aligned with the ensemble model’s final prediction. It highlights which edges consistently contribute to base learner decisions and are emphasized by the meta-learner’s attention mechanism.

\subsection{Evaluation Metrics}
To comprehensively assess both the classification performance and interpretability of the proposed framework, we employ the following evaluation metrics:

\paragraph{Accuracy}
Accuracy measures the proportion of correctly classified samples over the total number of samples:
\begin{equation}
\text{Accuracy} = \frac{TP + TN}{TP + TN + FP + FN}
\end{equation}
where \( TP \), \( TN \), \( FP \), and \( FN \) represent the number of true positives, true negatives, false positives, and false negatives, respectively.

\paragraph{Precision}
Precision evaluates the correctness of positive predictions, defined as:
\begin{equation}
\text{Precision} = \frac{TP}{TP + FP}
\end{equation}
A high precision indicates a low false positive rate.

\paragraph{Recall}
Recall (or sensitivity) assesses the model's ability to identify all relevant instances:
\begin{equation}
\text{Recall} = \frac{TP}{TP + FN}
\end{equation}
A high recall reflects a low false negative rate, which is particularly important in malware detection.

\paragraph{F1 Score}
The F1 Score is the harmonic mean of precision and recall, balancing the trade-off between the two:
\begin{equation}
\text{F1 Score} = 2 \cdot \frac{\text{Precision} \cdot \text{Recall}}{\text{Precision} + \text{Recall}}
\end{equation}

\paragraph{Fidelity}
Fidelity quantifies the explanatory power of a subgraph by assessing its influence on the model's prediction. It measures how the prediction outcome changes when either the important or unimportant portions of the graph are removed. Fidelity is evaluated in two complementary forms:

\begin{equation}
    Fidelity{+} = 1 - \frac{1}{N} \sum_{i=1}^{N} \mathds{1}\left( \hat{y}_i^{G_{C \setminus S}} = \hat{y}_i \right),
\end{equation}

\begin{equation}
    Fidelity{-} = 1 - \frac{1}{N} \sum_{i=1}^{N} \mathds{1}\left( \hat{y}_i^{G_{S}} = \hat{y}_i \right),
\end{equation}

where \( G_S \) denotes the important subgraph, \( G_C \) is the complete original graph, \( \hat{y}_i \) is the model's predicted label for the \( i \)-th sample, and \( y_i \) is the ground truth label.

$Fidelity{+}$ captures the impact of the removed important subgraph by comparing the model’s predictions on the original graph and on the graph with the important part removed (\( G_{C \setminus S} \)). A higher $Fidelity{+}$ indicates that the removed subgraph had a substantial influence on the model’s decision, thereby validating its importance in the prediction process.
Conversely, $Fidelity{-}$  assesses the predictive sufficiency of the important subgraph alone by comparing the model’s prediction on the full graph and on \( G_S \). A lower $Fidelity{-}$ implies that the identified subgraph retains most of the critical information needed for accurate prediction, thus indicating a more faithful and self-contained explanation.
Together, these two metrics provide a comprehensive evaluation of explanation quality by quantifying both the necessity and sufficiency of the identified subgraph in relation to the model’s output.

\section{Results and Analysis}
\label{sec:result}

In our experiments, we utilized malicious samples from the BODMAS~\cite{yang2021bodmas} and PMML~\cite{practicalsecurity2024pe} datasets, as well as benign samples obtained from the DikeDataset~\cite{dikedataset}. A summary of the datasets employed in this study is provided in Table~\ref{tab:dataset_stats}.
For dynamic CFG recovery, we leveraged the angr framework~\cite{shoshitaishvili2016state}, a Python-based binary analysis tool. Angr facilitates CFG construction by combining symbolic execution with constraint solving, allowing for precise and in-depth graph generation. In practice, angr disassembles binaries by lifting native x86-64 instructions into the VEX intermediate representation, which provides a uniform, architecture-independent form for analysis. Basic blocks are identified from the lifted IR and used as graph nodes, while directed edges are created to represent feasible control flow transfers between blocks. For binaries that employ packing or obfuscation, angr attempts to resolve indirect branches through symbolic execution.
All experiments were conducted on a workstation equipped with an Intel Xeon Platinum 8253 CPU (32 cores, 3.0 GHz) and 128 GB RAM. We implemented our framework in Python using PyTorch Geometric for model training and NetworkX v2.8.8 for graph processing and manipulation.

\begin{table}[h]
\centering
\caption{Characteristics of the evaluated datasets.}
\label{tab:dataset_stats}
\begin{tabular}{l*{4}{c}}
\toprule
\textbf{Dataset} & \textbf{\#Samples} & \textbf{Avg. Nodes} & \textbf{Avg. Edges} & \textbf{Label} \\
\midrule
BODMAS      & 122 & 63{,}226.44 & 66{,}033.72 & Malware \\
DikeDataset & 319 & 9{,}059.07  & 15{,}171.37 & Benign  \\
PMML        & 390 & 14{,}246.54 & 23{,}977.81 & Malware \\
\bottomrule
\end{tabular}%
\end{table}

To reduce the dimensionality of the initial 439-dimensional node feature vectors, we employed a symmetrical autoencoder that projects the data into a 64-dimensional latent space. The encoder consists of three fully connected layers with dimensions 439~$\rightarrow$256$\rightarrow$128$\rightarrow$64, each followed by a ReLU activation function. The decoder replicates this architecture in reverse, using layers of size 64$\rightarrow$128$\rightarrow$256$\rightarrow$~439, also with ReLU activations. The model was trained for 5000 epochs using the Adam optimizer with a learning rate of 0.0001, aiming to minimize the mean squared error (MSE) between the input features and their reconstructions. Training was terminated at 5000 epochs, as the validation MSE had stabilized below $1\times10^{-4}$ for the final 1000 epochs, indicating convergence. The resulting 64-dimensional representations from the encoder were then used as input node features for subsequent graph learning tasks.

\begin{figure*}[h]
    \centering
    \begin{subfigure}[b]{0.48\textwidth}
        \centering
        \includegraphics[width=\textwidth]{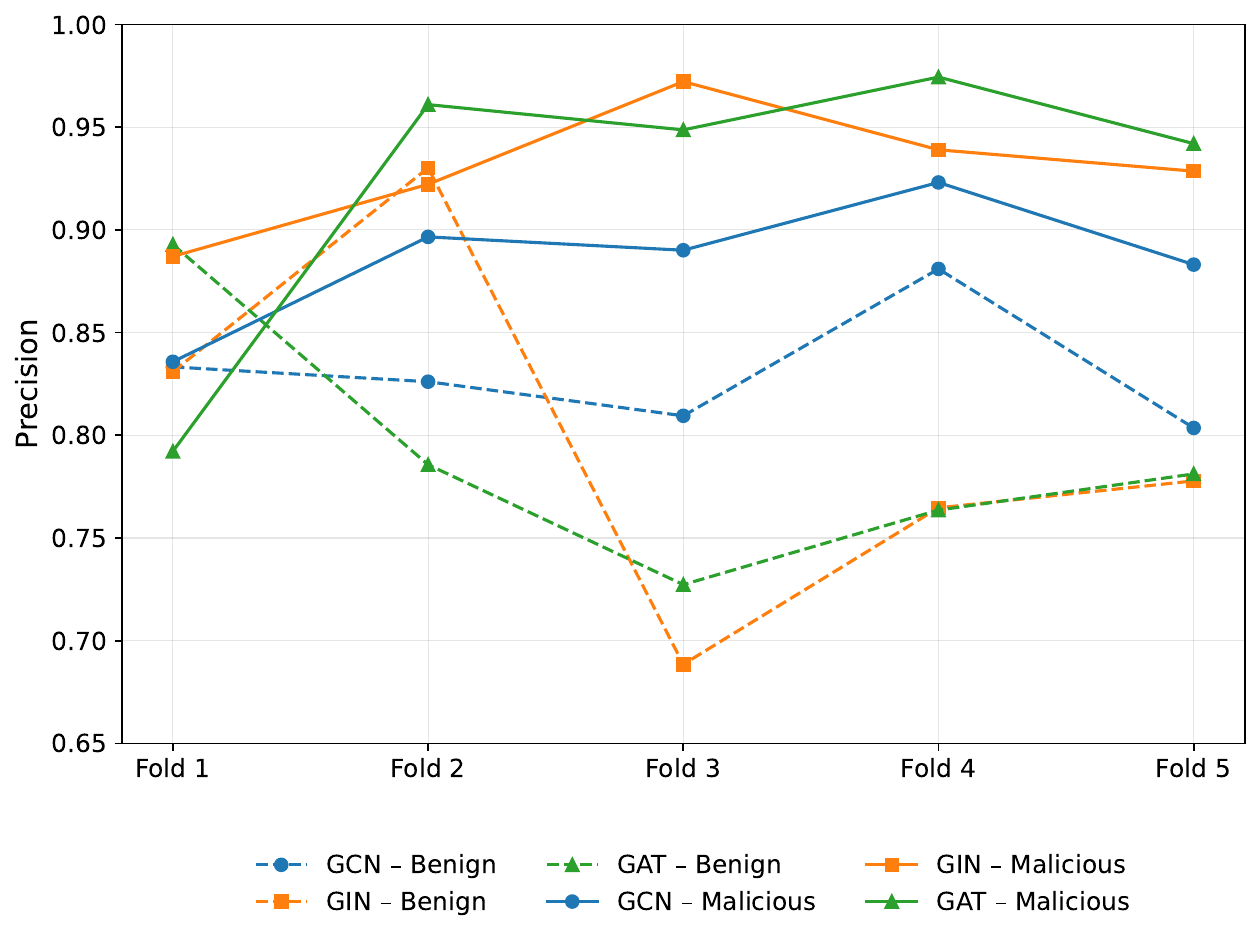}
        \label{fig:sub1}
    \end{subfigure}
    \hfill
    \begin{subfigure}[b]{0.48\textwidth}
        \centering
        \includegraphics[width=\textwidth]{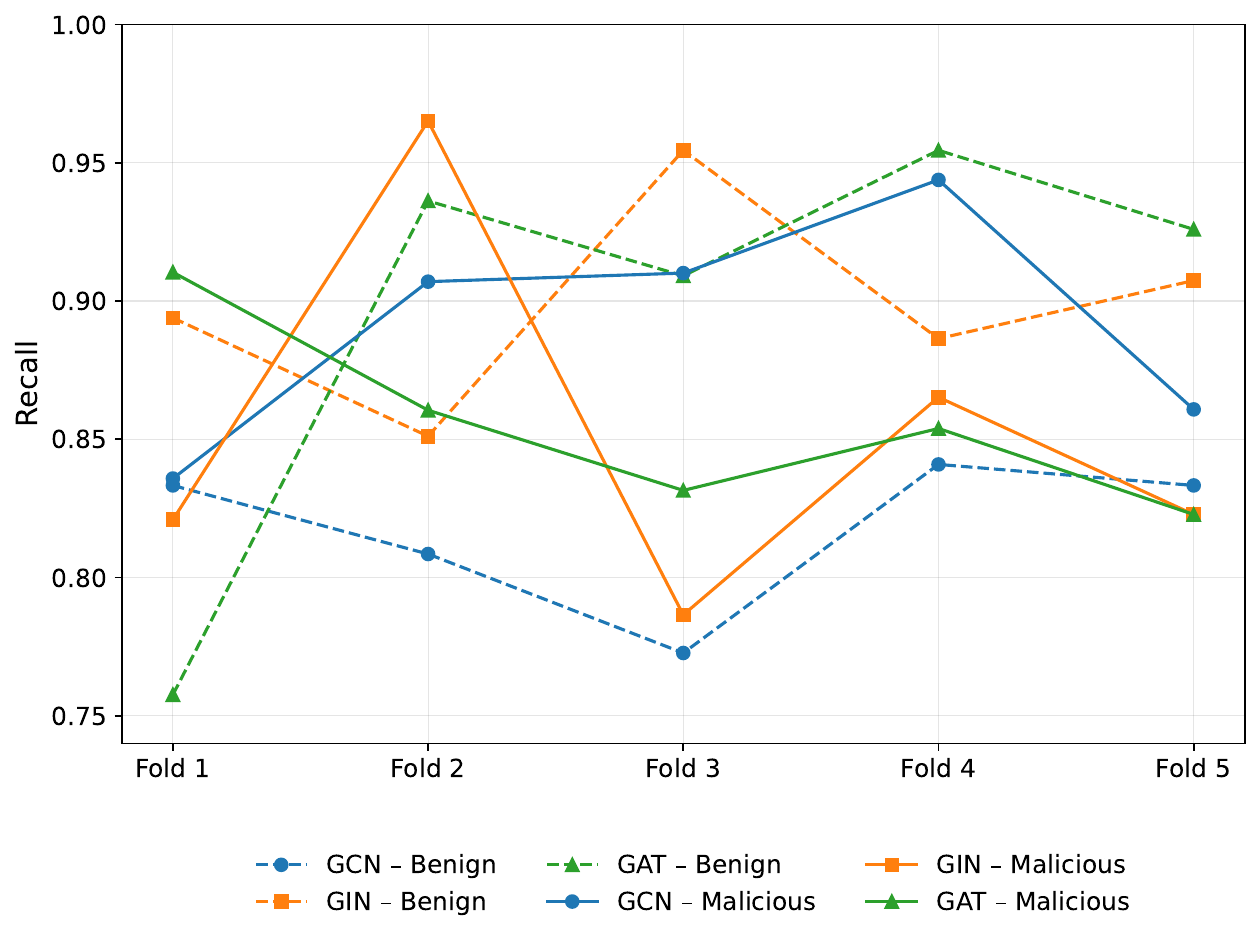}
        \label{fig:sub2}
    \end{subfigure}
    
    \vspace{0.25cm}
    
    \begin{subfigure}[b]{0.48\textwidth}
        \centering
        \includegraphics[width=\textwidth]{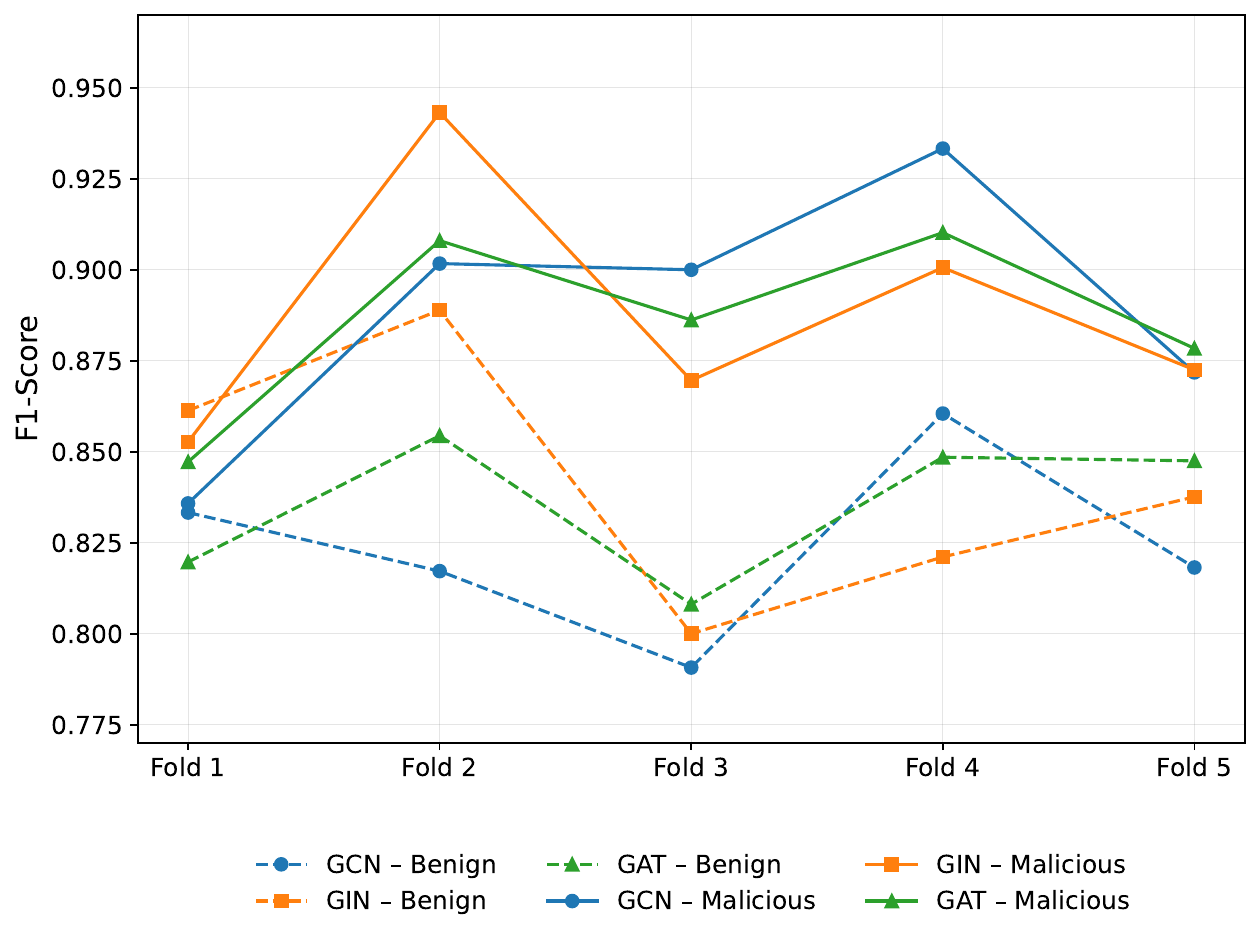}
        \label{fig:sub3}
    \end{subfigure}
    \hfill
    \begin{subfigure}[b]{0.48\textwidth}
        \centering
        \includegraphics[width=\textwidth]{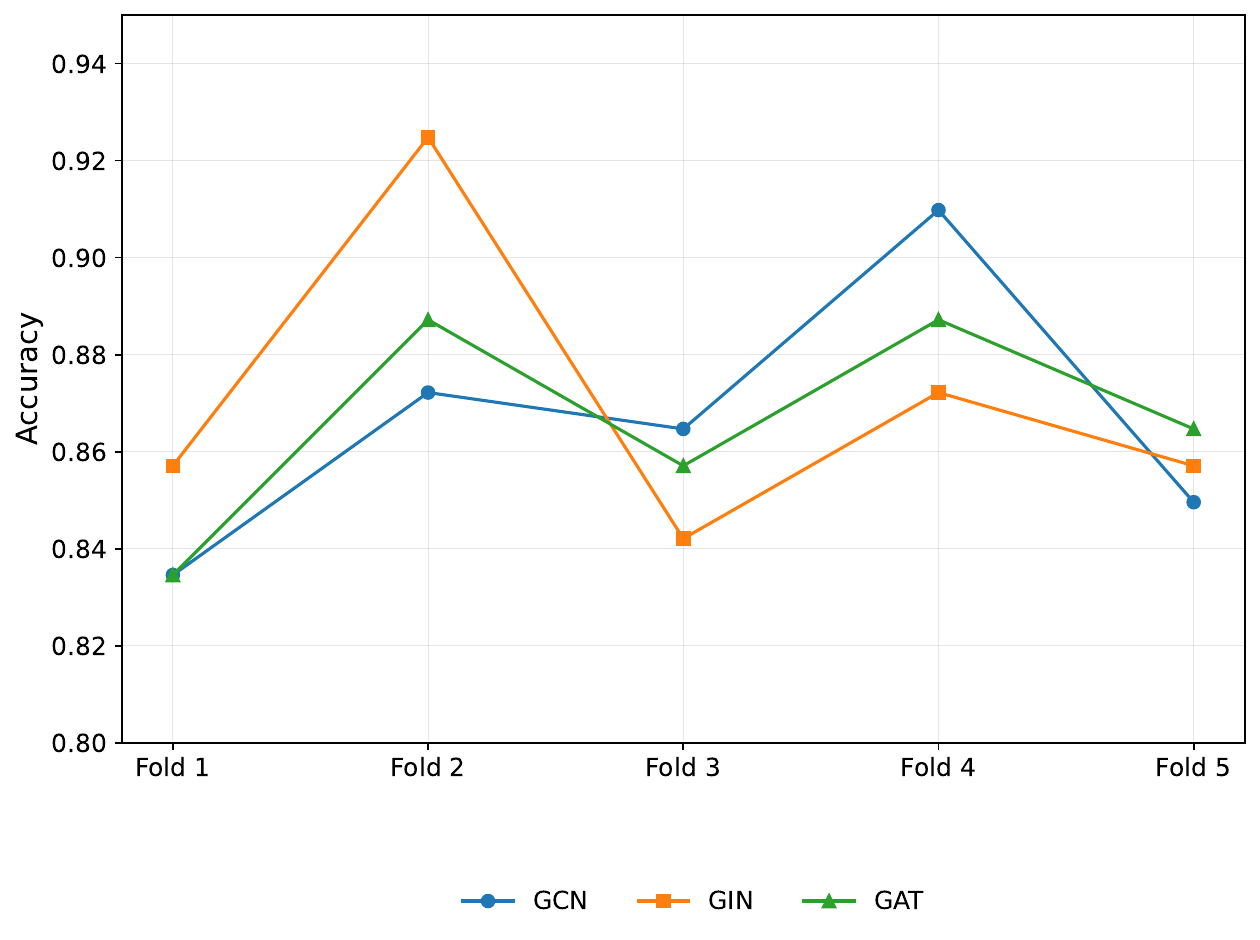}
        \label{fig:sub4}
    \end{subfigure}
    
    \caption{Validation performance metrics (Precision, Recall, F1-Score, and Accuracy) of the base learners (GCN, GAT, and GIN) across all five folds.}
    \label{fig:val_performance}
\end{figure*}
For the graph classification task, we experimented with three GNN architectures: GCN, GAT, and GIN, which served as base learners. The selection of GCN, GIN, and GAT as base learners was motivated by their complementary theoretical inductive biases for capturing distinct aspects of graph structure. The GCN operates on the principle of normalized neighborhood aggregation, which smooths features across local communities and provides a stable foundation for learning general topological roles \cite{GCN}. In contrast, the GIN is designed to be as powerful as the Weisfeiler-Lehman graph isomorphism test, making it exceptionally well-suited for discerning fine-grained topological differences and hierarchical structures \cite{GIN}. The GAT introduces a dynamic, feature-driven attention mechanism that adaptively weighs the importance of neighboring nodes, allowing it to focus on the most salient connections and resist structural noise \cite{GAT}. This trio was chosen because its members represent three distinct, non-overlapping paradigms (spectral smoothing, topological discrimination, and attention), thereby maximizing the diversity of structural perspectives for the ensemble.

All three base learners shared a common architecture comprising three graph convolutional layers with 64 hidden units each, followed by ReLU activation functions. A global mean pooling layer was applied to aggregate node-level embeddings into a graph-level representation. This representation was then passed through a dropout layer with a rate of 0.2, followed by a fully connected linear layer that produced class scores for binary classification.
Each base model was trained using the Adam optimizer with a learning rate of 0.001 and a weight decay of 0.0005. The training process used the cross-entropy loss function over 50 epochs. These hyperparameters, including learning rate, weight decay, and the number of training epochs, were selected through grid search.
Figure~\ref{fig:val_performance} presents the class-wise validation performance metrics of the base learners across all five folds.

After performing 5-fold cross-validation and generating new training data using the base learners’ predictions, all base models were retrained on the full training set to ensure consistency.
For the meta-learner, we employed an attention-based MLP. The architecture consists of three fully connected layers: an input layer of size equal to the concatenated outputs of the base learners, followed by two hidden layers with 128 and 64 units respectively, each with ReLU activations and dropout (rate 0.2), and a final output layer for binary classification. The meta-learner was trained for 100 epochs using the Adam optimizer with a learning rate of 0.001.

\begin{table}[h]
\centering
\caption{Performance comparison of individual base models (GCN, GIN, GAT), the Average Ensemble (AE), and the proposed Stacking Ensemble (SE) with an attention-based meta-learner.}
\label{tab:all_models_metrics_grouped}
\begin{tabular}{l|ccc|ccc|c}
\toprule
\multirow{2}{*}{\textbf{Model}} 
& \multicolumn{3}{c|}{\textbf{Benign}} 
& \multicolumn{3}{c|}{\textbf{Malicious}} 
& \multirow{2}{*}{\textbf{Accuracy}} \\
\cmidrule(lr){2-4} \cmidrule(lr){5-7}
& \textbf{Precision} & \textbf{Recall} & \textbf{F1 Score} 
& \textbf{Precision} & \textbf{Recall} & \textbf{F1 Score} 
&  \\
\midrule
GCN & 0.7903 & 0.7656 & 0.7778 & 0.8558 & 0.8725 & 0.8641 & 0.8313 \\
GIN & 0.7215 & \textbf{0.8906} & 0.7972 & \textbf{0.9195} & 0.7843 & 0.8466 & 0.8253 \\
GAT & 0.8364 & 0.7188 & 0.7731 & 0.8378 & 0.9118 & 0.8732 & 0.8373 \\
AE & 0.8036 & 0.7258 & 0.7627 & 0.8426&0.8922 &0.8667 & 0.8293 \\
SE  & \textbf{0.8475} & 0.7812 & \textbf{0.8130} & 0.8692 & \textbf{0.9118} & \textbf{0.8900} & \textbf{0.8614} \\
\bottomrule
\end{tabular}%
\end{table}

\begin{figure*}
    \centering
    \begin{subfigure}[b]{0.49\textwidth}
        \centering
        \includegraphics[width=\textwidth]{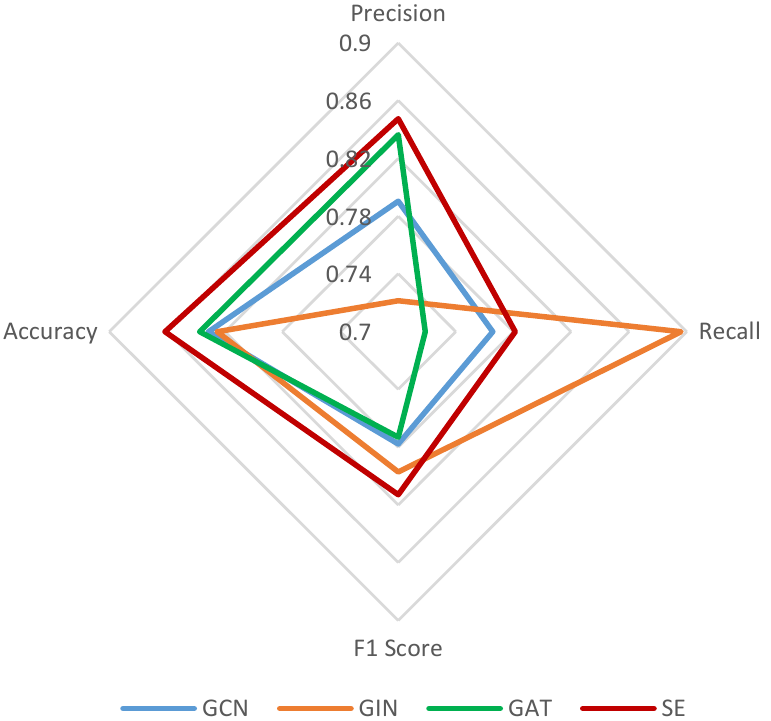}
        \caption{Benign class performance across models.}
        \label{fig:left}
    \end{subfigure}
    \hfill
    \begin{subfigure}[b]{0.49\textwidth}
        \centering
        \includegraphics[width=\textwidth]{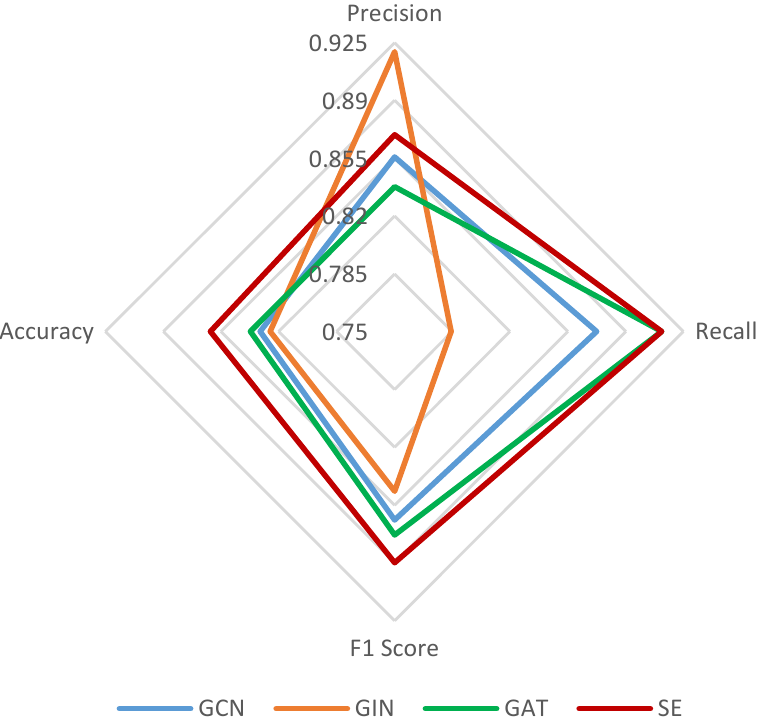}
        \caption{Malicious class performance across models.}
        \label{fig:right}
    \end{subfigure}
    
    \caption{Class-wise test performance.}
    \label{fig:radar}
\end{figure*}

To validate the effectiveness of the proposed SE method, we introduced a baseline Average Ensemble (AE) for comparison. In the AE approach, the softmax probabilities generated by the base GCN, GIN, and GAT models were averaged to produce the final prediction. As shown in Table~\ref{tab:all_models_metrics_grouped}, AE slightly improves the accuracy and F1 scores over individual base models, highlighting the benefit of ensemble aggregation. Nevertheless, the proposed SE achieves the best overall performance with an accuracy of 0.8614, surpassing AE by 3.2\%. This improvement confirms that the attention-based meta-learner in SE can effectively learn the relative importance of each base model, yielding a more adaptive and discriminative combination than the uniform weighting used in AE. Furthermore, Figure~\ref{fig:radar} provides a visual representation of these metrics using a radar plot.

Across all evaluation criteria, including precision, recall, F1-score, and accuracy, the SE model achieves the highest performance. This demonstrates that combining multiple base learners through ensemble learning enhances generalization.

The SE model achieves the highest overall accuracy of 86.14\%, along with the best macro and weighted averages across all metrics. For the malicious class, which is of primary concern in malware detection, the SE model obtains a recall of 91.18\%  and an F1-score of 89\%, outperforming each individual base learner. The high recall is particularly critical in cybersecurity applications, where false negatives, meaning undetected malware instances, are significantly more damaging than false positives.

Although the GIN model exhibits the highest precision for the malicious class (91.95\%), its recall is the lowest (78.43\%) among all models. This indicates that while GIN is effective in avoiding false positives, it is more prone to missing actual malicious instances. In malware detection, this trade-off is suboptimal, as failing to identify threats can lead to serious security breaches. Therefore, despite its strong precision, GIN's low recall diminishes its practical reliability as a standalone model.

In contrast, the SE model provides a balanced performance across both classes, achieving the highest F1-scores for benign and malicious samples. These findings underscore the advantage of ensemble strategies that integrate the strengths of diverse GNN architectures, resulting in a more effective and dependable detection framework.

\begin{figure}
    \centering
    \includegraphics[width=0.48\linewidth]{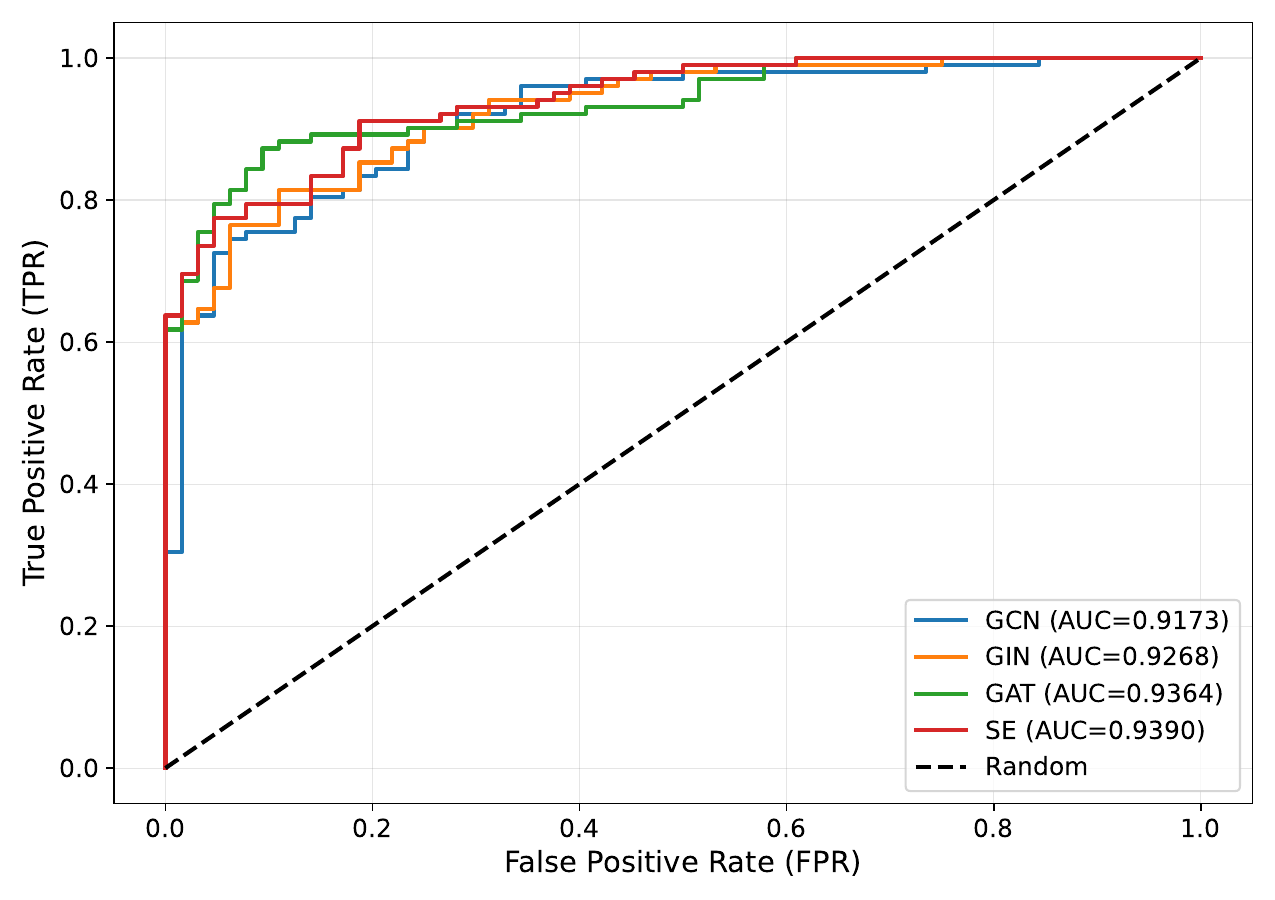}
    \caption{Comparison of ROC curves for base learners and SE model.}
    \label{fig:roc}
\end{figure}

Furthermore, Figure~\ref{fig:roc} presents the ROC curves for the base learners and the SE model. The Area Under the Curve (AUC) for the SE model reaches 0.9390, which is higher than that of all individual base models, further confirming the superior discriminative capability of the ensemble approach.

\begin{figure}
    \centering
    \includegraphics[width=0.55\linewidth]{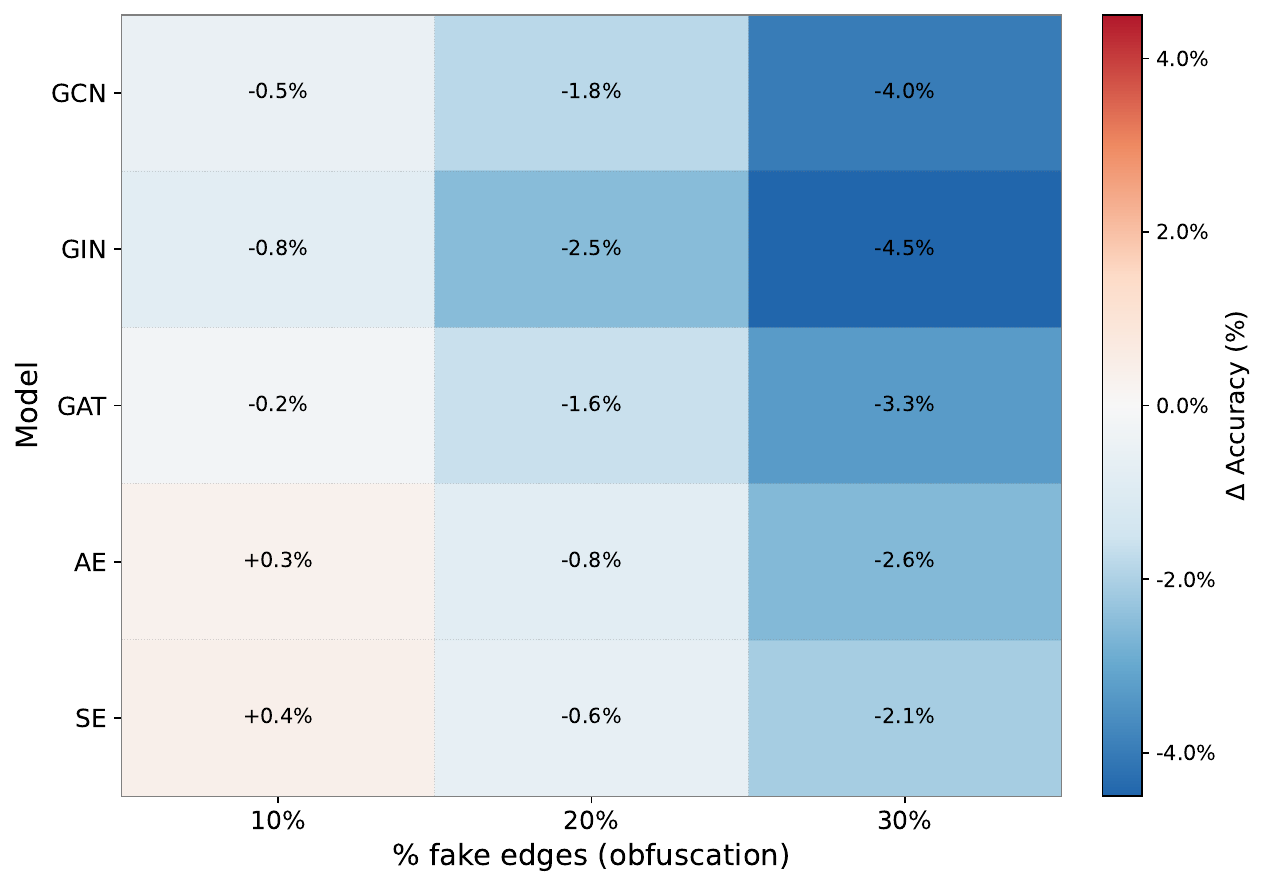}
    \caption{Accuracy variation under CFG obfuscation with 10–30\% fake-edge injection.}
    \label{fig:robustness}
\end{figure}

The robustness of the proposed framework against control flow obfuscation was examined through an adversarial test in which 10\%, 20\%, and 30\% fake edges were randomly injected into the CFGs of the test samples. The results are summarized in Table~\ref{tab:acc_obfuscation}, and the corresponding accuracy degradation is illustrated in Figure~\ref{fig:robustness}. As expected, the performance of all models decreases gradually with higher obfuscation levels; however, the proposed SE method exhibits the smallest decline (-2.1\% at 30\% obfuscation), outperforming both the base GNNs and the AE method. This demonstrates that SE is more resilient to structural perturbations, benefiting from the complementary graph representations learned by diverse GNN architectures.

\begin{table}[h]
\centering
\caption{Accuracy under CFG obfuscation via fake-edge injection.}
\label{tab:acc_obfuscation}
\begin{tabular}{l*{4}{c}}
\toprule
\textbf{Model} & \textbf{0\% Obf.} & \textbf{10\% Obf.} & \textbf{20\% Obf.} & \textbf{30\% Obf.} \\
\midrule
GCN & 0.8313 & 0.8263 & 0.8133 & 0.7913 \\
GIN & 0.8253 & 0.8173 & 0.8003 & 0.7803 \\
GAT & 0.8373 & 0.8353 & 0.8213 & 0.8043 \\
AE  & 0.8293 & 0.8323 & 0.8213 & 0.8033 \\
SE  & 0.8614 & 0.8654 & 0.8554 & 0.8404 \\
\bottomrule
\end{tabular}
\end{table}

\begin{figure*}[h]
    \centering
    \begin{subfigure}[b]{0.48\textwidth}
        \centering
        \includegraphics[width=\textwidth]{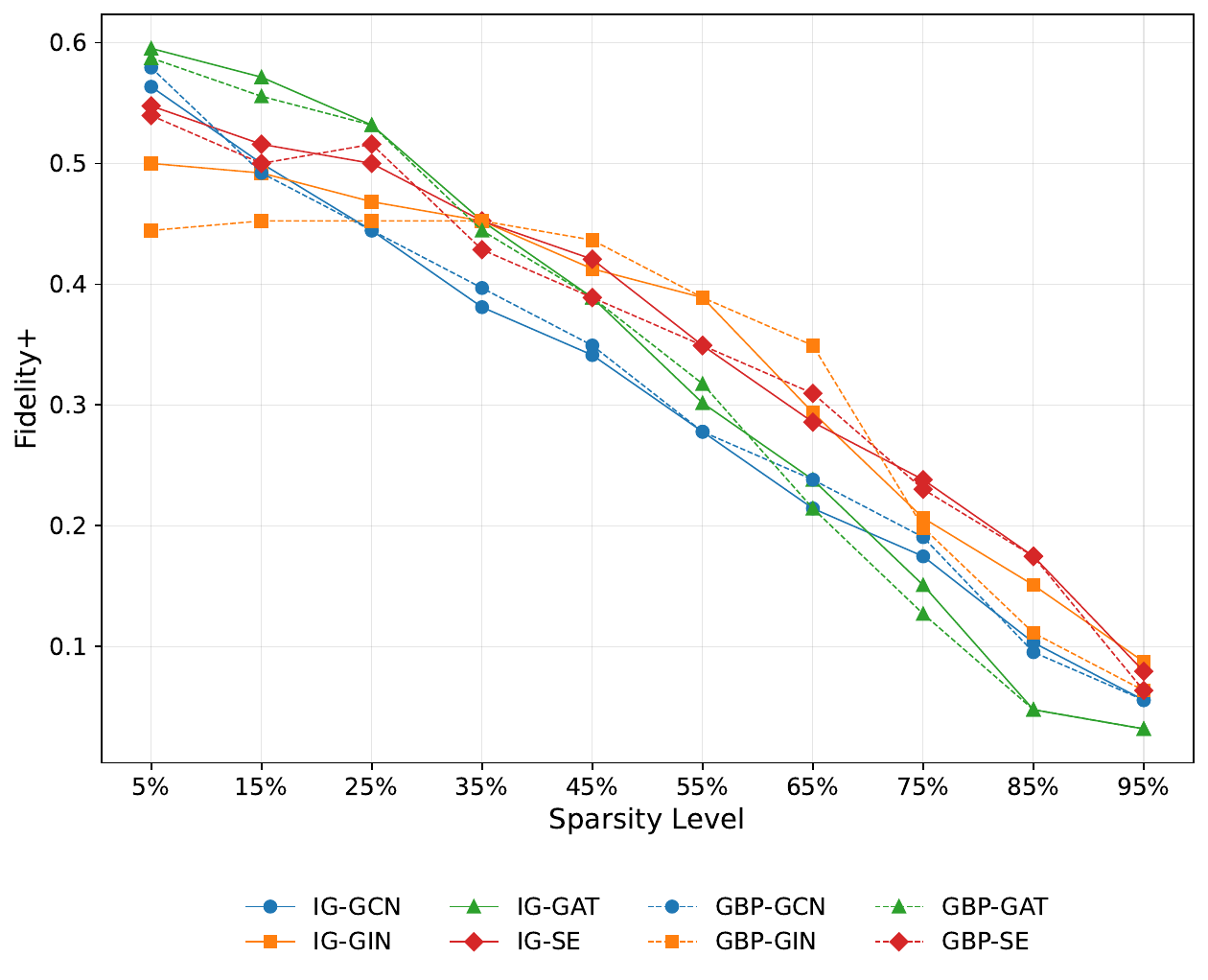}
        \label{fig:left}
    \end{subfigure}
    \hfill
    \begin{subfigure}[b]{0.48\textwidth}
        \centering
        \includegraphics[width=\textwidth]{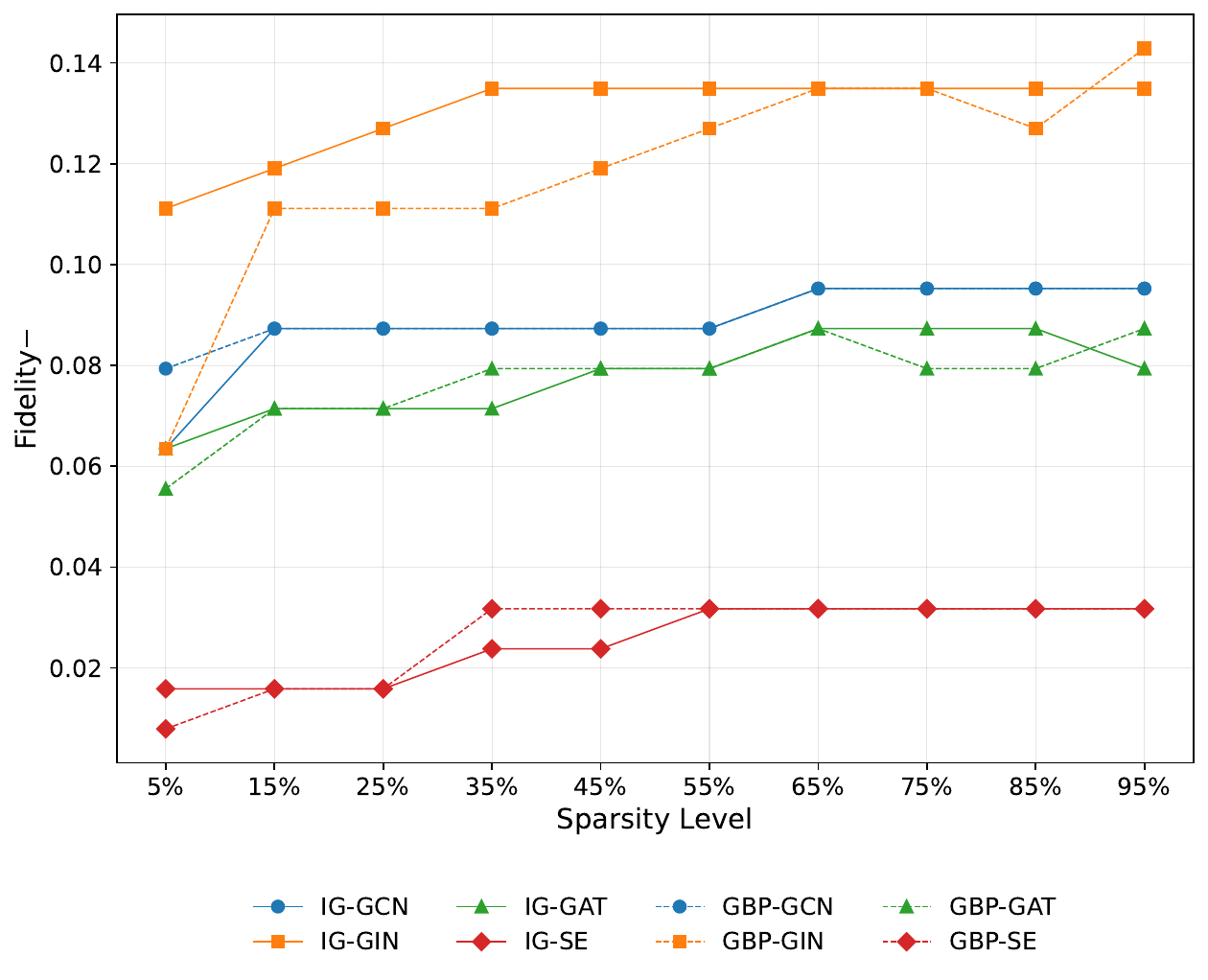}
        \label{fig:right}
    \end{subfigure}
    
    \caption{Fidelity evaluation of IG and GBP explainers across GCN, GIN, GAT, and SE models. For the SE model, IG and GBP explanations are further refined using the proposed aggregation-based method.}
    \label{fig:fidelity}
\end{figure*}

To evaluate explanation Fidelity, we employed Integrated Gradients (IG) and Guided Backpropagation (GBP) as representative post-hoc gradient-based explainers, both of which are widely regarded as state-of-the-art techniques for interpreting GNN predictions. For each base learner (GCN, GIN, and GAT), both IG and GBP were applied independently to generate explanations. In the case of the SE model, IG and GBP were also used as base explainers, and their outputs were further processed using the aggregation-based explanation method introduced in the previous section.
Figure~\ref{fig:fidelity} presents the Fidelity results for IG and GBP across the GCN, GIN, GAT, and SE models. The left subplot illustrates $Fidelity{+}$, while the right subplot depicts $Fidelity{-}$.

The results demonstrate that, for both IG and GBP, our aggregation-based explanation method applied to the SE model consistently achieves lower $Fidelity{-}$ values across different sparsity levels, indicating a stronger ability to identify and prioritize influential subgraphs. Furthermore, in terms of $Fidelity{+}$, the proposed explanation approach yields values that are comparable to those obtained from the base learners. This suggests that the aggregation method effectively preserves critical information while improving interpretability. Overall, these findings confirm the utility of the proposed explanation framework in the context of ensemble-based GNN models.

\subsection{Discussions}
Although the proposed framework quantitatively evaluates explanation fidelity, it is also important to consider the usability of the generated explanations for human analysts. In practice, explainable insights derived from the model (such as highlighted control flow regions or critical edges identified by the explainer) can help analysts trace malicious execution paths, verify suspicious behaviors, and prioritize samples for deeper manual inspection. These explanations provide a transparent mapping between model predictions and executable semantics, thereby facilitating trust and interpretability in real-world malware analysis workflows. While the current study focuses on quantitative evaluation, future work will involve human-centered validation through expert user studies to assess how effectively the generated explanations support analysts in reasoning, decision-making, and reducing investigation time.

Moreover, while the proposed framework achieves high detection accuracy and interpretability, its computational cost warrants discussion. The dynamic CFGs are generated using the angr-based environment, which naturally incurs higher processing time compared to static analysis. However, this additional cost is justified by the ability to capture runtime behaviors and obfuscated execution paths that significantly enhance detection reliability. In practice, CFG extraction can be parallelized across multiple analysis instances to improve throughput and support large-scale processing. In terms of deployment feasibility, the stacking ensemble engages multiple GNN base learners with distinct message-passing mechanisms, resulting in a moderate increase in computational demand during training and inference. Nevertheless, this design enhances generalization by combining complementary representations of program behavior. Overall, the additional computational cost is balanced by notable gains in accuracy, stability, and interpretability, making the framework suitable for scalable deployment in research and operational cybersecurity environments.

\section{Conclusion}\label{sec:conclusion}

This paper presented a novel SE framework for explainable malware detection using GNNs on dynamic CFGs. The proposed approach integrates multiple diverse GNN base learners with an attention-based meta-learner to enhance classification performance and provide interpretable predictions. A two-step node embedding strategy was employed to encode semantic and structural information from assembly instructions, while the meta-learner not only aggregated predictions but also offered insights into the relative contributions of each base model.

To address the challenge of interpretability in ensemble-based GNN models, we introduced a post-hoc explanation technique that fuses edge importance scores from gradient-based explainers using the attention weights of the meta-learner. This aggregation-driven approach generated ensemble-aware explanations aligned with the model’s final decision, improving the Fidelity.

Extensive experiments on real-world malware and benign datasets demonstrated that the proposed SE framework outperforms individual GNN models in terms of accuracy, F1-score, and AUC, particularly for the critical malicious class. Moreover, the explanation results revealed the capability of our method to identify influential subgraphs.

Future work may explore the extension of this framework to multi-class malware classification, integration of dynamic features beyond CFGs, and further investigation into explainability techniques tailored for ensemble architectures in graph-based learning contexts.


\bibliography{sn-bibliography}

\end{document}